\input jytex.tex   
\typesize=10pt
\magnification=1200
\baselineskip17truept
\hsize=6truein\vsize=8.5truein
\sectionnumstyle{blank}
\chapternumstyle{blank}
\chapternum=1
\sectionnum=1
\pagenum=0

\def\begintitle{\pagenumstyle{blank}\parindent=0pt\begin{narrow}[0.4in]}
\def\endtitle{\end{narrow}\newpage\pagenumstyle{arabic}}


\def\beginexercise{\vskip 20truept\parindent=0pt\begin{narrow}[10
truept]}
\def\endexercise{\vskip 10truept\end{narrow}}


\def\eql#1{\eqno\eqnlabel{#1}}
\def\ref{\reference}
\def\peq{\puteqn}
\def\pref{\putref}

\def\mgn{\marginnote}
\def\bex{\begin{exercise}}
\def\eex{\end{exercise}}


\font\open=msbm10 
\def\mbox#1{{\leavevmode\hbox{#1}}}

\def\hspace#1{{\phantom{\mbox#1}}}
\def\oZ{\mbox{\open\char90}}

\def\be{\beta}

\def\ka{\kappa}
\def\la{\lambda}
\def\La{\Lambda}
\def\om{\omega}

\def\th{\theta}

\def\ze{\zeta}

\def\De{\Delta}

\def\caL{{\cal L}}

\def\zf{$\zeta$--function}
\def\zfs{$\zeta$--functions}
\def\hk{heat-kernel}


\def\frac#1/#2{\leavevmode\kern.1em
\raise.5ex\hbox{\the\scriptfont0 #1}\kern-.1em/\kern-.15em
\lower.25ex\hbox{\the\scriptfont0 #2}}
\def\sfrac#1/#2{\leavevmode\kern.1em
\raise.5ex\hbox{\the\scriptscriptfont0 #1}\kern-.1em/\kern-.15em
\lower.25ex\hbox{\the\scriptscriptfont0 #2}}

\def\gtorder{\mathrel{\raise.3ex\hbox{$>$}\mkern-14mu
             \lower0.6ex\hbox{$\sim$}}}
\def\ltorder{\mathrel{\raise.3ex\hbox{$<$}\mkern-14mu
             \lower0.6ex\hbox{$\sim$}}}

\def\semidirprod{\rlap{\ss C}\raise1pt\hbox{$\mkern.75mu\times$}}
\def\for{\lower6pt\hbox{$\Big|$}}
\def\fish{\kern-.25em{\phantom{abcde}\over \phantom{abcde}}\kern-.25em}


\def\boxit#1{\vbox{\hrule\hbox{\vrule\kern3pt
        \vbox{\kern3pt#1\kern3pt}\kern3pt\vrule}\hrule}}
\def\dalemb#1#2{{\vbox{\hrule height .#2pt
        \hbox{\vrule width.#2pt height#1pt \kern#1pt
                \vrule width.#2pt}
        \hrule height.#2pt}}}
\def\square{\mathord{\dalemb{5.9}{6}\hbox{\hskip1pt}}}

\def\frac#1#2{{{#1}\over{#2}}}


\def\etc{{\it etc. }}

\def\eg{{\it e.g. }}
\def\ie{{\it i.e. }}
\def\cf{{\it cf }}
\def\pa{\partial}


  %

\def\3j#1#2#3#4#5#6{\left\lgroup\matrix{#1&#2&#3\cr#4&#5&#6\cr}
\right\rgroup}

\def\man{{\cal M}}
\def\caI{{\cal I}}

\def\m?{\mgn{?}}

\def\pa{\partial}

\def\beq{\begin{eqnarray}}
\def\eeq{\end{eqnarray}}


\def\aop#1#2#3{{\it Ann. Phys.} {\bf {#1}} (19{#2}) #3}

\def\cmp#1#2#3{{\it Comm. Math. Phys.} {\bf {#1}} (19{#2}) #3}
\def\cqg#1#2#3{{\it Class. Quant. Grav.} {\bf {#1}} (19{#2}) #3}

\def\jgp#1#2#3{{\it J. Geom. and Phys.} {\bf {#1}} (19{#2}) #3}
\def\jmp#1#2#3{{\it J. Math. Phys.} {\bf {#1}} (19{#2}) #3}
\def\jpa#1#2#3{{\it J. Phys.} {\bf A{#1}} (19{#2}) #3}

\def\np#1#2#3{{\it Nucl. Phys.} {\bf B{#1}} (19{#2}) #3}
\def\pl#1#2#3{{\it Phys. Lett.} {\bf {#1}} (19{#2}) #3}

\def\prD#1#2#3{{\it Phys. Rev.} {\bf D{#1}} (19{#2}) #3}

\def\cras#1#2#3{{\it Comptes Rend. Acad. Sci. (Paris)} {\bf{#1}} (#2) #3}

\def\mpcps#1#2#3{{\it Math. Proc. Camb. Phil. Soc.} {\bf{#1}} (19{#2}) #3}

\def\am#1#2#3{{\it Acta Mathematica} {\bf {#1}} (19{#2}) #3}
\def\aim#1#2#3{{\it Adv. in Math.} {\bf {#1}} (19{#2}) #3}
\def\ajm#1#2#3{{\it Am. J. Math.} {\bf {#1}} ({#2}) #3}

\def\aom#1#2#3{{\it Ann. of Math.} {\bf {#1}} (19{#2}) #3}

\def\cpde#1#2#3{{\it Comm. Partial Diff. Equns.} {\bf {#1}} (19{#2}) #3}

\def\invm#1#2#3{{\it Invent. Math.} {\bf {#1}} (19{#2}) #3}
\def\ijpam#1#2#3{{\it Ind. J. Pure and Appl. Math.} {\bf {#1}} (19{#2}) #3}
\def\jdg#1#2#3{{\it J. Diff. Geom.} {\bf {#1}} (19{#2}) #3}
\def\jfa#1#2#3{{\it J. Func. Anal.} {\bf {#1}} (19{#2}) #3}

\def\jmpa#1#2#3{{\it J. Math. Pures. Appl.} {\bf {#1}} ({#2}) #3}

\def\ojm#1#2#3{{\it Osaka J.Math.} {\bf {#1}} ({#2}) #3}

\def\pja#1#2#3{{\it Proc. Jap. Acad.} {\bf {A#1}} (19{#2}) #3}

\def\tams#1#2#3{{\it Trans. Am. Math. Soc.} {\bf {#1}} (19{#2}) #3}

\begin{title}
\vglue 1truein
\vskip15truept
\centertext {\Bigfonts \bf The $N \cup D$ problem} \vskip 20truept
\centertext{J.S.Dowker\footnote{dowker@a13.ph.man.ac.uk}} \vskip 7truept
\centertext{\it Department of Theoretical Physics, }
\centertext{\it The University of Manchester,}
\centertext{ \it Manchester, England} \vskip 20truept \centertext {Abstract}
\vskip10truept
\begin{narrow}
The hybrid spectral problem where the field satisfies Dirichlet conditions (D) 
on part of the boundary of the relevant domain and Neumann (N) on the
remainder is discussed in simple terms. A conjecture for the $C_1$
coefficient is presented and the conformal determinant on a 2-disc, where
the $D$ and $N$ regions are semi-circles, is derived.
Comments on higher coefficients are made.
\end{narrow}
\vskip 5truept
\vskip 60truept
\vfil
\end{title}
\pagenum=0
\section{\bf1. Introduction}
The explicit construction of the general form of the \hk\ expansion
coefficients has reached the stage when further progress is impeded mainly
by ungainliness. Unless there is some compelling reason for finding a
specific higher coefficient, its exhibition is not particularly
enlightening and is not really worth the, often considerable, effort.
Other, more productive, avenues consist of generalising the differential
operator, the manifold or the boundary conditions. In the latter context a
simply stated extension is the class of problems where the field satisfies
Dirichlet conditions (D), say, on part of the boundary and Neumann (N)
on the remainder. These boundary conditions are sometimes termed `mixed '
in the classical literature (\eg Sneddon [\pref{Sneddon}]) or sometimes
`hybrid' (\eg Treves [\pref{treves2}] chap.37).

It is anticipated that the \hk\ coefficients will receive
contributions from the codimension-2 junction of these two regions. This
has been confirmed by Avramidi [\pref{avramidi}] and work by  van
den Berg and Gilkey, [\pref{VdBandG}], on heat {\it content} is also pertinent.

In this work we wish to make some observations on this question that are
mainly example driven and with a minimum of algebra. It is hoped that these
considerations will prove useful in more general field and string theoretic
areas where \hk\ coefficients play important roles in divergence and
scaling questions.

\section{\bf2. Basic idea}

A simple calculation, or the drawing of a few modes, shows that on the
interval of length $L$ with Dirichlet (D) and Neumann (N) conditions, the
various problems are related by

$$\eqalign{ (D,N)_L\cup(D,D)_L&=(D,D)_{2L}\cr
(D,N)_L\cup(N,N)_L&=(N,N)_{2L}\cr} \eql{relns}$$ $$
(D,D)_L\cup(N,N)_L=P_{2L} \eql{relns2}$$ where the notation $(D,N)$
signifies a problem with $D$ conditions at one end and $N$ at the other and
$P$ stands for periodic conditions. Averaging (\peq{relns}) gives, using
(\peq{relns2}) $$ (D,N)\cup{1\over2}P_{2L}={1\over2}P_{4L}\,.
\eql{relns3}$$

The `subtraction' implied by (\peq{relns}) and (\peq{relns3}), in order
to extract the $(D,N)$ part, amounts to a cull of the even modes on the
doubled interval, as is well known (\cf Rayleigh [\pref{Rayleigh}],
vol I, p.247).

These relations can be applied to the arc of a circle, which might form
part of an SO(2) foliation of a two--dimensional region (or the projection
of a higher dimensional region onto two dimensions). A wedge is a good example
which we will now look at. Say the angle of the wedge is $\be$, then the
relations (\peq{relns}) apply equally well, where the notation means that
either $D$ or $N$ applies on the straight sides of the wedge, (say $\th=0$ and
$\th=\be$). Equation (\peq{relns}) can be immediately applied to the
heat-kernel and its small-time expansion to determine the form of the
heat-kernel coefficients in the $(D,N)$ combination. We will show how this
works out for the $C_1$ coefficient.  The $(D,D)$ and $(N,N)$ wedge
coefficients are well known,

$$ C^{\rm wedge}_1(D,D)=C^{\rm wedge}_1(N,N)={\pi^2-\be^2\over6\be}.
\eql{cee1s}$$
Hence from (\peq{relns})
$$ C^{\rm wedge}_1(D,N)=-{\pi^2+2\be^2\over12\be}.
\eql{cee1DN}$$
This result has been derived by Watson in a rather complicated way using
the modes directly, [\pref{Watson}].

Sommerfeld, [\pref{Sommerfeld}] vol.2 p.827, also mentions the `mixed' wedge
and indicates how to treat it using images if $\be=\pi n/m$.

Incidentally the conjecture by Gottlieb (equn.(3.5) in [\pref{Gottlieb}]),
that the $(N,D)$ case differs from the $(D,D)$ one only by a sign, is
incorrect, although it is true in the special case of a right-angled
wedge, as is easily checked by looking at rectilinear flat domains.
\section{\bf3. The general case}
Consider now in general dimension a manifold whose boundary is piecewise
smooth  consisting of domains $\pa\man_i$ which intersect in codimension-2
manifolds, ${\cal I}_{ij}$. On each of the pieces $\pa\man_i$ either $D$ or
$N$ is imposed. Then, by Kac's principle, $C_1$ will take contributions
from the manifolds of codimension zero, one and two independently.

In general dimension, for all $D$ or all $N$, the smeared coefficients are
known

$$ C_1(D)=\bigg({1\over6}-\xi\bigg)\int Rf dV+\int\bigg({1\over3}\ka -
{1\over2}(n.\pa)\bigg)f\,dS+{1\over6}\int{\pi^2-\be^2\over\be}\,f\,dL$$

$$ C_1(N)=\bigg({1\over6}-\xi\bigg)\int R dV+\int\bigg({1\over3}\ka
-2\psi+
{1\over2}(n.\pa)\bigg)f\,dS+{1\over6}\int{\pi^2-\be^2\over\be}\,f\,dL$$
where the integrals over $S$ and $L$ symbolically include summations over
$i$ and $(i,j)$. Neumann conditions have been extended to Robin in
these formulae.

For a mixture of $D$ and $N$, the volume contribution clearly remains
unchanged while the surface contribution divides simply into a sum
separately over those regions $\pa\man(D)$ and $\pa\man(N)$ subject to $D$
and $N$ respectively. The codimension 2 intersections ${\cal I}_{ij}$
divide into the three (wedge) types ${\cal I}(D,D)$, ${\cal I}(N,N)$ and
${\cal I}(N,D)$. So our conjecture for the corresponding $C_1$ is

$$ \eqalign{C_1(D,N)&=\bigg({1\over6}-\xi\bigg)\int_\man Rf
dV+\int_{\pa\man(D)}\bigg({1\over3}\ka
-{1\over2}(n.\pa)\bigg)f\,dS+\cr&\int_{\pa\man(N)}\bigg({1\over3}\ka
-2\psi+ {1\over2}(n.\pa)\bigg)f\,dS+{1\over6}\int_{\cal I(D,D)\cup\cal
I(N,N)} {\pi^2-\be^2\over\be}\,f\,dL-\cr &{1\over12}\int_{\cal
I(D,N)}{\pi^2+2\be^2\over\be}f\,dL, \cr}\eql{cee1}$$
where we have, perhaps cavalierly, extended $N$ to Robin. Dimensions
show that the
codimension-2 contribution cannot depend on the boundary function, $\psi$.

If the boundary is smooth, then all the dihedral angles $\be$ equal $\pi$
and the codimension-2 part of (\peq{cee1}) reduces to

$$ -{\pi\over4}\int_{\cal I(D,N)}f\,dL
\eql{cee12}$$

For example, for the 3-ball with $D$ on the northern hemisphere and
$N\,(\psi=0)$ on the southern, $$ C_1(D,N)={8\pi\over3}-{\pi^2\over2}$$ for
$f=1$.

A local derivation of (\peq{cee12}), justifying the application of Kac's
principle, has been given by Avramidi [\pref{avramidi}]. It has also been
obtained by van den Berg (unpublished).
\section{\bf4. The lune}
Relations (\peq{relns}), (\peq{relns2}) can also be applied to the 2-lune
where the intervals are the sections of the lines of latitude cut out by the
the two longitudes, $\phi=0$, $\phi=\be$.
In this case the extrinsic curvatures vanish (the boundaries are geodesically
embedded) but there is a volume (area) term independent of the
boundary conditions.

The \zfs\ are now somewhat more explicit [\pref{D2,ChandD}]. It is possible to
work with general angle $\be$. If we choose $\be=\pi/q$, $q\in\oZ$, the \zfs\
have been derived in [\pref{ChandD}] and used in [\pref{AandD}].

Denoting the lune
by $\caL(\be)$ we have
$$\eqalign{
(D,N)_{\caL(\be)}\cup(D,D)_{\caL(\be)}&=(D,D)_{\caL(2\be)}\cr
(D,N)_{\caL(\be)}\cup(N,N)_{\caL(\be)}&=(N,N)_{\caL(2\be)}\cr}
$$
so that the corresponding \zfs\ combine algebraically,
$$
\ze_\be^{ND}(s)=\ze_{2\be}^{DD}(s)-\ze_\be^{DD}(s)
=\ze_{2\be}^{NN}(s)-\ze_\be^{NN}(s)\,.
\eql{ndzeta}$$

The $DD$ and $NN$
\zfs\ have been derived in [\pref{ChandD}] as Barnes \zfs\ for conformal
coupling in three dimensions (leading to simple eigenvalues) and yield the
specific value, for example,
$$
\ze_\be^{DD}(0)={1\over12}\bigg({\pi\over\be}-{\be\over2\pi}\bigg),
$$
which can be used to confirm the expression (\peq{cee1DN}) using the relation
between $C_1$ and $\ze(0)$. (In this case there are no zero modes.)

The volume contribution, $\be/6$, to $C_1$ is standard and is the same for all
boundary conditions. Hence the contribution of each $(N,D)$ corner
(of which there are two) is
$$
{1\over2}\bigg[-{4\pi\over24}\bigg({\pi\over\be}+{\be\over \pi}\bigg)
-{\be\over6}\bigg]=
-{\pi^2+2\be^2\over12\be}
$$
as required.
\section{\bf5. The disc and semi-circle}
The fact that the extrinsic curvatures are zero means that the lune is not
excessively helpful in deriving the form of the higher coefficients in the
$(D,N)$ case. Some further information can be obtained by looking at the
half-disc  with semi-circular boundary having different conditions on the
diameter and circumference.

A straightforward application of, say the Stewartson and Waechter Laplace
transform
technique combined with an image method soon yields the results for the
short time expansions
$$
K_{DD}(t)\sim{1\over8t}-{2+\pi\over8\sqrt{\pi t}}+{5\over24}
+{\sqrt t(\pi+16)\over256
\sqrt\pi}+\bigg({1\over315}+{1\over32}\bigg)\,t+\ldots
\eql{ssDD}$$
$$
K_{ND}(t)\sim{1\over8t}+{2-\pi\over8\sqrt{\pi t}}-{1\over24}
+{\sqrt t(\pi-16)\over256
\sqrt\pi}+\bigg({1\over315}-{1\over32}\bigg)\,t+\ldots
\eql{ssND}$$
$$
K_{NN}(t)\sim{1\over8t}+{2+\pi\over8\sqrt{\pi t}}+{5\over24}
+{\sqrt t(5\pi+48)\over256
\sqrt\pi}+\bigg({1\over45}+{3\over32}\bigg)\,t+\ldots
\eql{ssNN}$$
$$
K_{DN}(t)\sim{1\over8t}-{2-\pi\over8\sqrt{\pi t}}-{1\over24}
+{\sqrt t(5\pi-48)\over256
\sqrt\pi}+\bigg({1\over45}-{3\over32}\bigg)\,t+\ldots
\eql{ssDN}$$
where $DN$ means $D$ on the diameter and $N$ on the circumference, \etc

The constant terms check with (\peq{cee1s}) and (\peq{cee1DN}) for
$\be=\pi/2$. Also (\peq{relns2}), applied to the diameter as a wedge of
angle $\pi$, yields the $D$ and $N$ (\eg [\pref{Moss}]), {\it full} disc
expansions.

The extrinsic curvature vanishes on the diameter and equals one on the
circumference
part of the boundary so some information on the $C_{3/2}$ and $C_2$
coefficients
can be inferred. Formulae in the non-mixed types $(D,D)$ and $(N,N)$ have been
given in [\pref{AandD,DandA}] which agree with the relevant parts of the above
expressions. Indeed we used the hemi-disc in deriving these results.

 Also in [\pref{AandD}] will be found an expression for $C_2$ in the case
the boundary parts $\pa\man_i$ are subject to Robin conditions with different
boundary functions, $\psi_i$ although all dihedral angles are restricted to
$\pi/2$.

In the case of $C_2$, the $1/315$ is the contribution of the curved $D$
semicircle while the $\pm1/32$ is the effect of the (two) corners and likewise
regarding the $1/45\pm3/32$ combination. The $C_{3/2}$ coefficient exhibits
a similar
structure. Experience with the flat wedge shows that
it is unwise to draw too many conclusions when the angle is $\pi/2$. What we
can say,
however, is that, using the $3/2$ coefficient as an exemplar, one term will
have the general form
$$\eqalign{
-{\sqrt\pi\over24}\bigg[\int_{\caI(D,D)}\la_{DD}&(\be)(\ka_1+\ka_2)+
\int_{\caI(N,N)}\la_{NN}(\be)(\ka_1+\ka_2)
+\cr
&\int_{\caI(N,D)}\big(\la_{ND}(\be)\ka_D+\la_{DN}(\be)\ka_N\big)\bigg]\cr}
$$
where $\la_{ND}(\pi/2)=-\la_{DD}(\pi/2)=3$ and $\la_{DN}(\pi/2)=
-\la_{NN}(\pi/2)=-9$.
This change of sign is a simple consequence of images, or of (\peq{relns})
since the
$DD$ and $NN$ quantities vanish when $\be=\pi$.
\section{\bf6. The disc determinant}
A direct attack via modes, of what is, after rectilinear domains, the
simplest two-dimensional situation, \ie
a disc subject to $N$ on one half of the circumference, and $D$ on the rest,
would
seem to be difficult in so far as the construction of the \zf\ or heat-kernel
is
concerned. However, the functional determinant appears to be accessible by
conformal transformation from that on an $ND$-lune of angle $\pi$, \ie a
hemisphere with $N$ on one half of the rim (the equator) and $D$ on the rest,
which is an easy quantity to find in terms of Barnes \zf\ from
(\peq{ndzeta}). For this to work, one
would need the conjectured form of $C_1$, (\peq{cee1}), to be valid in order to
construct the required cocycle function in two dimensions. Applying the
standard techniques this is (\cf [\pref{D}]), for a {\it smooth} boundary,
$$\eqalign{
W[e^{-2\om} g,g]={1\over24\pi}\int_\man\om\big(R+\square\om\big)\,dV
+{1\over12\pi}
\int_{\pa\man}\om\big(\ka+{1\over2}(n.\pa)\om\big)\,dS+\cr
{1\over8\pi}\bigg(\int_{\pa\man(N)}-\int_{\pa\man(D)}\bigg)(n.\pa)\,\om\,dS
-{1\over16}\sum_k\om_k\cr}
\eql{cocycle}$$
where $k$ labels the points where $D$ and $N$ meet and $\om_k$ are the values
of $\om$ at these points. If $\pa\man(D)$ is empty there is a volume term
coming from the pure $N$ zero mode.

To go from the hemisphere to the disc we employ the equatorial stereographic
projection as in [\pref{Weis,BandG,AandD,DandA,D}] noting that there is no
codimension-2
contribution because the conformal factor is unity on the boundary, implying
$\om_k=0$.

Then (\peq{cocycle}) can be written
$$
W_{ND}[\bar g,g]={1\over2}\bigg(W_{D}[\bar g,g]
+\overline W_{N}[\bar g,g]\bigg),
\eql{cocycle2}$$
where $\overline W_{N}$ means the usual Neumann expression, omitting the
zero mode
piece, and we can use the known values, ($\bar g$ = disc and $g$ = hemisphere),
$$\eqalign{
W_D[\bar g,g]&={1\over6}\log2-{1\over3}\cr
\overline W_N[\bar g,g]&={2\over3}\log2+{1\over6}.\cr}
\eql{cocycle3}$$

The \zf\ on the $ND$-hemisphere follows from (\peq{ndzeta}) with $\be=\pi$.
The \zf, $\ze_\pi^{DD}(s)$ is the usual hemisphere \zf\ and the determinant has
been considered a number of times.
$\ze_{2\pi}^{DD}(s)$, corresponds to Sommerfeld's double covering of
three-space introduced in connection with the half-plane boundary.

Since one needs conformal invariance in two dimensions, not three, the
\zfs\ are actually {\it modified} Barnes \zfs\ which have been dealt with
in [\pref{D2,Dow1,Cook}]. The determinants can be computed generally in
terms of Barnes \zfs\ but, because of the rational nature of $\pi/\be$, in
this case, they can be reduced to Epstein or Hurwitz \zfs. The
general theory, appropriate to the arbitrary 2-lune, is developed in
[\pref{D2}]. However it is probably easier to proceed directly.

From [\pref{D2}] the \zf\ for $-\De$ on the $ND$ 2-hemisphere is
$$
\ze^{ND}_\pi=\ze^{DD}_{2\pi}(s)-\ze^{DD}_\pi
=\sum_{m,n=0}^\infty{1\over\big((1+m+n)^2-1/4\big)^s}
$$

Expanding in the $1/4$ leads to the expression for the derivative at 0,
$$
{\ze^{ND}_{\pi}}'(0)=\ze_2'(0,1/2\mid1,1)+\ze_2'(0,3/2\mid1,1)-{N_2(1)\over4}
\eql{deriv1}$$
where
$$
\ze_2(s,a\mid1,1)=\sum_{m,n=0}^\infty{1\over(a+m+n)^s}
$$
is a 2-dimensional Barnes \zf\ and $N_2(a)$ is its residue
at $s=2$; $N_2(a)=1$.

In this simple case the sums can easily be rearranged,
$$
\sum_{m,n=0}^\infty{1\over(a+m+n)^s}=\sum_{N=0}^\infty{N+1\over(N+a)^s}=
\ze_R(s-1,a)+(1-a)\,\ze_R(s,a)
$$
so that
$$
\ze_2(s,1/2\mid1,1)+\ze_2(s,3/2\mid1,1)=2\ze_R(s-1,1/2)=2(2^{s-1}-1)\,
\ze_R(s-1)
$$
and therefore from (\peq{deriv1})
$$
{\ze^{ND}_{\pi}}'(0)=-\ze_R'(-1)-{1\over12}\log2-{1\over4}.
\eql{NDdet}$$
The absence of a $\ze_R'(0)$ term is related to the absence of the perimeter
\hk\ coefficient caused by the equal--sized $N$ and $D$ regions.
The \zf\ has only the Weyl volume pole.

For comparison the standard formulae for the $DD$ and $NN$-hemispheres are
$$
{\ze^{DD}_\pi}'(0)=2\ze_R'(-1)-\ze_R'(0)-{1\over4}
$$
and
$$
{\ze^{NN}_\pi}'(0)=2\ze_R'(-1)+\ze_R'(0)-{1\over4}.
$$
By conformal transformation, on the $ND$-disc, our final result is
$$
W_{ND}^{\rm disc}={1\over2}\ze_R'(-1)+{11\over24}\log2-{1\over24}
$$
using (\peq{cocycle2}) with (\peq{cocycle3}) and defining $W=-\ze'(0)/2$ by
convention.
\section{\bf 6. Conclusion}

Apart from rectilinear domains, and the hemisphere, there seem few
situations that can be solved exactly for $ND$-conditions and this is a
drawback to the construction of the precise forms of the \hk\ coefficients.
Nevertheless we have made a certain progress in a simple minded way making
use of the $ND$-wedge expression. This type of reasoning can be extended to
higher dimensions leading to information about the trihedral corner
contributions and their higher analogues. Surprisingly the conformal
functional determinant is available on the `half-N half-D' disc by
conformal transformation from the $ND$-hemisphere, and has been computed,
assuming the conjecture for the coefficient $C_1$ is correct.

\section{\bf References}
\vskip 5truept
\begin{putreferences}
\ref{AandD}{Apps,J.S. and Dowker,J.S. \cqg{15}{98}{1121}.}
\ref{APS}{Atiyah,M.F., V.K.Patodi and I.M.Singer: Spectral asymmetry and
Riemannian geometry \mpcps{77}{75}{43}.} \ref{avramidi}{Avramidi,I. {\it
Heat kernel asymptotics of a non-smooth boundary value problem} Workshop on
Spectral Geometry, Bristol 2000.} \ref{AandT}{Awada,M.A. and D.J.Toms:
Induced gravitational and gauge-field actions from quantised matter fields
in non-abelian Kaluza-Klein thory \np{245}{84}{161}.} \ref{BandI}{Baacke,J.
and Y.Igarishi: Casimir energy of confined massive quarks
\prD{27}{83}{460}.} \ref{Barnesa}{Barnes,E.W.: On the Theory of the
multiple Gamma function {\it Trans. Camb. Phil. Soc.} {\bf 19} (1903) 374.}
\ref{Barnesb}{Barnes,E.W.: On the asymptotic expansion of integral
functions of multiple linear sequence, {\it Trans. Camb. Phil. Soc.} {\bf
19} (1903) 426.} \ref{Barv}{Barvinsky,A.O. Yu.A.Kamenshchik and
I.P.Karmazin: One-loop quantum cosmology \aop {219}{92}{201}.}
\ref{BandM}{Beers,B.L. and Millman, R.S. :The spectra of the
Laplace-Beltrami operator on compact, semisimple Lie groups.
\ajm{99}{1975}{801-807}.} \ref{BandH}{Bender,C.M. and P.Hays: Zero point
energy of fields in a confined volume \prD{14}{76}{2622}.} \ref{BBG}{Bla\v
zi\' c,N., Bokan,N. and Gilkey,P.B.: Spectral geometry of the form valued
Laplacian for manifolds with boundary \ijpam{23}{92}{103-120}}
\ref{BEK}{Bordag,M., E.Elizalde and K.Kirsten: { Heat kernel coefficients
of the Laplace operator on the D-dimensional ball}, \jmp{37}{96}{895}.}
\ref{BGKE}{Bordag,M., B.Geyer, K.Kirsten and E.Elizalde,: { Zeta function
determinant of the Laplace operator on the D-dimensional ball},
\cmp{179}{96}{215}.} \ref{BKD}{Bordag,M., K.Kirsten,K. and Dowker,J.S.:
Heat kernels and functional determinants on the generalized cone
\cmp{182}{96}{371}.} \ref{Branson}{Branson,T.P.: Conformally covariant
equations on differential forms \cpde{7}{82}{393-431}.}
\ref{BandG2}{Branson,T.P. and Gilkey,P.B. {\it Comm. Partial Diff. Eqns.}
{\bf 15} (1990) 245.} \ref{BandG}{Branson,T.P. and Gilkey,P.B.
\tams{344}{94}{479}.} \ref{BGV}{Branson,T.P., P.B.Gilkey and
D.V.Vassilevich {\it The Asymptotics of the Laplacian on a manifold with
boundary} II, hep-th/9504029.} \ref{BCZ1}{Bytsenko,A.A, Cognola,G. and
Zerbini, S. : Quantum fields in hyperbolic space-times with finite spatial
volume, hep-th/9605209.} \ref{BCZ2}{Bytsenko,A.A, Cognola,G. and Zerbini,
S. : Determinant of Laplacian on a non-compact 3-dimensional hyperbolic
manifold with finite volume, hep-th /9608089.} \ref{CandH2}{Camporesi,R.
and Higuchi, A.: Plancherel measure for $p$-forms in real hyperbolic space,
\jgp{15}{94}{57-94}.} \ref{CandH}{Camporesi,R. and A.Higuchi {\it On the
eigenfunctions of the Dirac operator on spheres and real hyperbolic
spaces}, gr-qc/9505009.} \ref{ChandD}{Chang, Peter and Dowker,J.S.
\np{395}{93}{407}.} \ref{cheeg1}{Cheeger, J.: Spectral Geometry of Singular
Riemannian Spaces. \jdg {18}{83}{575}.} \ref{cheeg2}{Cheeger,J.: Hodge
theory of complex cones {\it Ast\'erisque} {\bf 101-102}(1983) 118-134}
\ref{Chou}{Chou,A.W.: The Dirac operator on spaces with conical
singularities and positive scalar curvature, \tams{289}{85}{1-40}.}
\ref{CandT}{Copeland,E. and Toms,D.J.: Quantized antisymmetric tensor
fields and self-consistent dimensional reduction in higher-dimensional
spacetimes, \break\np{255}{85}{201}} \ref{Cook}{Cook,A. 1986 {\it PhD
Thesis}, University of Manchester.} \ref{DandH}{D'Eath,P.D. and
J.J.Halliwell: Fermions in quantum cosmology \prD{35}{87}{1100}.}
\ref{cheeg3}{Cheeger,J.:Analytic torsion and the heat equation. \aom{109}
{79}{259-322}.} \ref{DandE}{D'Eath,P.D. and G.V.M.Esposito: Local boundary
conditions for Dirac operator and one-loop quantum cosmology
\prD{43}{91}{3234}.} \ref{DandE2}{D'Eath,P.D. and G.V.M.Esposito: Spectral
boundary conditions in one-loop quantum cosmology \prD{44}{91}{1713}.}
\ref{D}{Dowker,J.S. \cqg{11}{94}{557}.} \ref{Dow1}{Dowker,J.S.
\cmp{162}{94} {633}.} \ref{Dow8}{Dowker,J.S. {\it Robin conditions on the
Euclidean ball} MUTP/95/7; hep-th\break/9506042. {\it Class. Quant.Grav.}
to be published.} \ref{Dow9}{Dowker,J.S. {\it Oddball determinants}
MUTP/95/12; hep-th/9507096.} \ref{Dow10}{Dowker,J.S. {\it Spin on the
4-ball}, hep-th/9508082, {\it Phys. Lett. B}, to be published.}
\ref{DandA}{Dowker,J.S. and Apps,J.A. \cqg{12}{95}{1363}.}
\ref{D2}{Dowker,J.S. \jmp{35}{94}{4989}; 1995 (Feb.) erratum.}
\ref{DandA2}{Dowker,J.S. and J.S.Apps, {\it Functional determinants on
certain domains}. To appear in the Proceedings of the 6th Moscow Quantum
Gravity Seminar held in Moscow, June 1995; hep-th/9506204.}
\ref{DABK}{Dowker,J.S., Apps,J.S., Bordag,M. and Kirsten,K.: Spectral
invariants for the Dirac equation with various boundary conditions {\it
Class. Quant.Grav.} to be published, hep-th/9511060.}
\ref{EandR}{E.Elizalde and A.Romeo : An integral involving the generalized
zeta function, {\it International J. of Math. and Phys.} {\bf13} (1994)
453.} \ref{ELV2}{Elizalde, E., Lygren, M. and Vassilevich, D.V. : Zeta
function for the laplace operator acting on forms in a ball with gauge
boundary conditions. hep-th/9605026} \ref{ELV1}{Elizalde, E., Lygren, M.
and Vassilevich, D.V. : Antisymmetric tensor fields on spheres: functional
determinants and non-local counterterms, \jmp{}{96}{} to appear. hep-th/
9602113} \ref{Kam2}{Esposito,G., A.Y.Kamenshchik, I.V.Mishakov and
G.Pollifrone: Gravitons in one-loop quantum cosmology \prD{50}{94}{6329};
\prD{52}{95}{3457}.} \ref{Erdelyi}{A.Erdelyi,W.Magnus,F.Oberhettinger and
F.G.Tricomi {\it Higher Transcendental Functions} Vol.I McGraw-Hill, New
York, 1953.} \ref{Esposito}{Esposito,G.: { Quantum Gravity, Quantum
Cosmology and Lorentzian Geometries}, Lecture Notes in Physics, Monographs,
Vol. m12, Springer-Verlag, Berlin 1994.} \ref{Esposito2}{Esposito,G. {\it
Nonlocal properties in Euclidean Quantum Gravity}. To appear in Proceedings
of 3rd. Workshop on Quantum Field Theory under the Influence of External
Conditions, Leipzig, September 1995; gr-qc/9508056.} \ref{EKMP}{Esposito G,
Kamenshchik Yu A, Mishakov I V and Pollifrone G.: One-loop Amplitudes in
Euclidean quantum gravity. \prd {52}{96}{3457}.} \ref{ETP}{Esposito,G.,
H.A.Morales-T\'ecotl and L.O.Pimentel {\it Essential self-adjointness in
one-loop quantum cosmology}, gr-qc/9510020.} \ref{FORW}{Forgacs,P.,
L.O'Raifeartaigh and A.Wipf: Scattering theory, U(1) anomaly and index
theorems for compact and non-compact manifolds \np{293}{87}{559}.}
\ref{GandM}{Gallot S. and Meyer,D. : Op\'erateur de coubure et Laplacian
des formes diff\'eren-\break tielles d'une vari\'et\'e riemannienne
\jmpa{54}{1975} {289}.} \ref{Gilkey1}{Gilkey, P.B, Invariance theory, the
heat equation and the Atiyah-Singer index theorem, 2nd. Edn., CRC Press,
Boca Raton 1995.} \ref{Gilkey2}{Gilkey,P.B.:On the index of geometric
operators for Riemannian manifolds with boundary \aim{102}{93}{129}.}
\ref{Gilkey3}{Gilkey,P.B.: The boundary integrand in the formula for the
signature and Euler characteristic of a manifold with boundary
\aim{15}{75}{334}.} \ref{Gottlieb}{Gottlieb, H.P.W., {\it J.Aust.Math.Ass.}
{\bf26} (1984) 293.} \ref{Grubb}{Grubb,G. {\it Comm. Partial Diff. Eqns.}
{\bf 17} (1992) 2031.} \ref{GandS1}{Grubb,G. and R.T.Seeley
\cras{317}{1993}{1124}; \invm{121}{95} {481}.} \ref{GandS}{G\"unther,P. and
Schimming,R.:Curvature and spectrum of compact Riemannian manifolds,
\jdg{12}{77}{599-618}.} \ref{IandT}{Ikeda,A. and Taniguchi,Y.:Spectra and
eigenforms of the Laplacian on $S^n$ and $P^n(C)$.
\ojm{15}{1978}{515-546}.} \ref{IandK}{Iwasaki,I. and Katase,K. :On the
spectra of Laplace operator on $\La^*(S^n)$ \pja{55}{79}{141}.}
\ref{JandK}{Jaroszewicz,T. and P.S.Kurzepa: Polyakov spin factors and
Laplacians on homogeneous spaces \aop{213}{92}{135}.}
\ref{Kam}{Kamenshchik,Yu.A. and I.V.Mishakov: Fermions in one-loop quantum
cosmology \prD{47}{93}{1380}.} \ref{KandM}{Kamenshchik,Yu.A. and
I.V.Mishakov: Zeta function technique for quantum cosmology {\it Int. J.
Mod. Phys.} {\bf A7} (1992) 3265.} \ref{KandC}{Kirsten,K. and Cognola.G,: {
Heat-kernel coefficients and functional determinants for higher spin fields
on the ball} \cqg{13}{96} {633-644}.} \ref{Levitin}{Levitin,M.: { Dirichlet
and Neumann invariants for Euclidean balls}, {\it Diff. Geom. and its
Appl.}, to be published.} \ref{Luck}{Luckock,H.C.: Mixed boundary
conditions in quantum field theory \jmp{32}{91}{1755}.}
\ref{MandL}{Luckock,H.C. and Moss,I.G,: The quantum geometry of random
surfaces and spinning strings \cqg{6}{89}{1993}.} \ref{Ma}{Ma,Z.Q.: Axial
anomaly and index theorem for a two-dimensional disc with boundary
\jpa{19}{86}{L317}.} \ref{Mcav}{McAvity,D.M.: Heat-kernel asymptotics for
mixed boundary conditions \cqg{9}{92}{1983}.} \ref{MandV}{Marachevsky,V.N.
and D.V.Vassilevich {\it Diffeomorphism invariant eigenvalue \break problem
for metric perturbations in a bounded region}, SPbU-IP-95, \break
gr-qc/9509051.} \ref{Milton}{Milton,K.A.: Zero point energy of confined
fermions \prD{22}{80}{1444}.} \ref{MandS}{Mishchenko,A.V. and Yu.A.Sitenko:
Spectral boundary conditions and index theorem for two-dimensional
manifolds with boundary \aop{218}{92}{199}.} \ref{Moss}{Moss,I.G.
\cqg{6}{89}{759}.} \ref{MandP}{Moss,I.G. and S.J.Poletti: Conformal anomaly
on an Einstein space with boundary \pl{B333}{94}{326}.}
\ref{MandP2}{Moss,I.G. and S.J.Poletti \np{341}{90}{155}.}
\ref{NandOC}{Nash, C. and O'Connor,D.J.: Determinants of Laplacians, the
Ray-Singer torsion on lens spaces and the Riemann zeta function
\jmp{36}{95}{1462}.} \ref{NandS}{Niemi,A.J. and G.W.Semenoff: Index theorem
on open infinite manifolds \np {269}{86}{131}.} \ref{NandT}{Ninomiya,M. and
C.I.Tan: Axial anomaly and index thorem for manifolds with boundary
\np{245}{85}{199}.} \ref{norlund2}{N\"orlund~N. E.:M\'emoire sur les
polynomes de Bernoulli. \am {4}{21} {1922}.} \ref{Poletti}{Poletti,S.J.
\pl{B249}{90}{355}.} \ref{RandT}{Russell,I.H. and Toms D.J.: Vacuum energy
for massive forms in $R^m\times S^N$, \cqg{4}{86}{1357}.}
\ref{Rayleigh}{Rayleigh,Lord, {\it The Theory of Sound}, Vols. I and II,
2nd Edn., (MacMillan, London, 1894).} \ref{RandS}{R\"omer,H. and
P.B.Schroer \pl{21}{77}{182}.} \ref{Schulman}{Schulman.L.S.}
\ref{Sneddon}{Sneddon,I.N.,{\it Mixed Boundary Value Problems in Potential
Theory}, (North-Holland, Amsterdam, 1966).}
\ref{Sommerfeld}{Sommerfeld,A. in {\it Die Differential- und Integralgleichungen
der Mechanik und Physik} by Frank,P. and v.Mises,R., (Vieweg, Braunschweig, 
1935).}
\ref{Trautman}{Trautman,A.: Spinors and Dirac operators on hypersurfaces
\jmp{33}{92}{4011}.} \ref{VdBandG}{van den Berg,M. and Gilkey,P. \jfa
{120}{94}{48}.} \ref{Vass}{Vassilevich,D.V.{Vector fields on a disk with
mixed boundary conditions} gr-qc /9404052.} \ref{Voros}{Voros,A.: Spectral
functions, special functions and the Selberg zeta function.
\cmp{110}{87}439.} \ref{Watson}{Watson,S. 1998 {\it PhD Thesis.},
University of Bristol.} \ref{Weis}{Weisberger,W.I. \cmp {112}{87}{633}.}
\ref{Ray}{Ray,D.B.: Reidemeister torsion and the Laplacian on lens spaces
\aim{4}{70}{109}.} \ref{McandO}{McAvity,D.M. and Osborn,H. Asymptotic
expansion of the heat kernel for generalised boundary conditions
\cqg{8}{91}{1445}.} \ref{AandE}{Avramidi,I. and Esposito,G. Heat kernel
asymptotics with generalised boundary conditions, hep-th/9701018.}
\ref{MandS}{Moss,I.G. and Silva P.J., Invariant boundary conditions for
gauge theories gr-qc/9610023.}
\ref{barv}{Barvinsky,A.O.\pl{195B}{87}{344}.} \ref{krantz}{Krantz,S.G.
Partial Differential Equations and Complex Analysis (CRC Press, Boca Raton,
1992).} \ref{treves}{Treves,F. Introduction to Pseudodifferential and
Fourier Integral Operators,\break Vol.1, (Plenum Press,New York,1980).}
\ref{treves2}{Treves,F. {\it Basic linear partial differential equations}
(Academic Press, New York, 1975).} \ref{EandS}{Egorov,Yu.V. and Shubin,M.A.
Partial Differential Equations (Springer-Verlag, Berlin,1991).}
\ref{AandS}{Abramowitz,M. and Stegun,I.A. Handbook of Mathematical
Functions (Dover, New York, 1972).} \ref{ACNY}{Abouelsaood,A., Callan,C.G.,
Nappi,C.R. and Yost,S.A.\np{280}{87} {599}.} \ref{BGKE}{Bordag,M., B.Geyer,
K.Kirsten and E.Elizalde, { Zeta function determinant of the Laplace
operator on the D-dimensional ball}, \cmp{179}{96}{215}.}

\end{putreferences}
\bye